\documentclass{article}

\newcommand{\ket}[1]{|#1\rangle}
\newcommand{\bra}[1]{\langle #1 |}

\title{Quantum multiparty key distribution protocol without use
of entanglement}
\author{Ryutaroh Matsumoto\\
Dept.\ of Communications and Integrated Systems\\
Tokyo Institute of Technology, 152-8550 Japan}
\date{August 7, 2007}

\begin{document}
\maketitle
\begin{abstract}
We propose a quantum key distribution (QKD) protocol
that enables three parties agree at once on a shared common random
bit string in presence of an eavesdropper without use of
entanglement. We prove its unconditional security and analyze
the key rate.
\end{abstract}

\section{Introduction}
When a sender, Alice,
 wants to send a confidential message to single receiver, Bob,
over a public communication channel in an unconditionally
secure way, they have to share a secret common random bit string,
which is usually called a secret key.
Such a secret key can be shared by a quantum key distribution
(QKD) protocol, such as the BB84 protocol \cite{bennett84,shor00}
even when there is an eavesdropper, Eve, with unlimited computational
power.

When Alice sends a confidential message by broadcast to two receivers,
Bob and Charlie, these three parties have to share a common secret key.
We propose a protocol enabling them to share such a secret key
under the assumption that there are point-to-point quantum channels
from Alice to Bob and Charlie whose message
can be eavesdropped and modified by Eve and three point-to-point
classical channels among Alice, Bob and Charlie
whose message can be eavesdropped but cannot be modified by Eve.
Eve is assumed to do whatever manipulation on quantum systems
transmitted over the quantum channels allowed by the quantum mechanics.
This assumption is the same as Ref.~\cite{chen05}
and is a multiparty generalization of that
in Ref.~\cite{shor00}.
This problem is called the multiparty key distribution or
conference key agreement.
Under this assumption we prove the unconditional security and
give a lower bound on the key rate.

As a prior relevant research,
it was pointed out in Ref.~\cite{singh03} that
Alice can secretly send a key after two different secret keys
are shared between Alice and Bob and between Alice and Charlie
by a conventional two party QKD protocol such as Ref.~\cite{bennett84}.
The difference of our proposed protocol to Ref.~\cite{singh03}
is that our protocol allows three parties to share a secret key
at once.

As another prior relevant research,
Chen and Lo \cite{chen05} proposed a protocol for the same goal
as our proposed protocol. In their protocol,
Alice has to prepare the Bell state while in our protocol
she does not need an entangled quantum state,
which makes our protocol easier to implement with current
technology than theirs \cite{chen05}.
Our protocol is the first multiparty QKD protocol that does
not use entangled state.
It is also worth noting that
the security proof for our protocol does not use
multipartite entanglement distillation \cite{murano98},
while Chen and Lo \cite{chen05} use it.

This paper is organized as follows:
Section 2 presents the proposed protocol.
Section 3 shows an unconditional security proof and
a lower bound on its key rate.
Section 4 gives concluding remarks.

\section{Protocol}
In this section we describe our proposed protocol.
Let $\ket{0}$, $\ket{1}$ be an orthonormal basis
for a qubit state space, and
$\ket{+} = (\ket{0}+\ket{1})/\sqrt{2}$,
$\ket{-} = (\ket{0}+\ket{1})/\sqrt{2}$.
We also define the matrices $X$ and $Z$ representing
the bit error and the phase error, respectively, as
\begin{eqnarray*}
&&X\ket{0} = \ket{1},\quad, X\ket{1}=\ket{0},\\
&&Z\ket{+} = \ket{-},\quad, Z\ket{-}=\ket{+}.
\end{eqnarray*}

\begin{enumerate}
\item\label{l1}
Alice makes a random qubit sequence according to
the i.i.d.\ uniform distribution on $\{\ket{0}$, $\ket{1}$,
$\ket{+}$, $\ket{-}\}$ and sends it to Bob.
Alice also sends the same qubit sequence to Charlie.
\item\label{l2}
Bob chooses the $\{\ket{0}$, $\ket{1}\}$ basis or $\{\ket{+}$, $\ket{-}\}$
basis uniformly randomly for each received qubit and measure it
by the chosen basis.
\item\label{l3}
Charlie does the same thing as Step~\ref{l2}.
\item\label{l4}
Alice publicly announces which basis $\{\ket{0}$, $\ket{1}\}$ or
$\{\ket{+}$, $\ket{-}\}$ each transmitted qubit belongs to.
Bob and Charlie also publicly announce which bases were used for measurement
of each qubit.
In the following steps they will only consider qubits with which
transmission basis and measuring bases  coincide among all of them.
\item\label{l5}
Suppose that there are $2n$ qubits transmitted in the $\{\ket{0}$, $\ket{1}\}$
basis and measured with the $\{\ket{0}$, $\ket{1}\}$ basis by both Bob and Charlie.
Index those qubits by $1$, \ldots, $2n$.
Define the bit $a_i = 0$ if Alice's $i$-th qubit was $\ket{0}$,
and $a_i=1$ otherwise.
Define the bit $b_i = 0$ if Bob's measurement outcome
for $i$-th qubit was $\ket{0}$,
and $b_i=1$ otherwise.
Define the bit $c_i = 0$ if Charlie's measurement outcome
for $i$-th qubit was $\ket{0}$,
and $c_i=1$ otherwise.
\item\label{l6}
Suppose also that there are $2n'$ qubits transmitted in the $\{\ket{+}$, $\ket{-}\}$
basis and measured with the $\{\ket{+}$, $\ket{-}\}$ basis by both Bob and Charlie.
Index those qubits by $1$, \ldots, $2n'$.
Define the bit $\alpha_i = 0$ if Alice's $i$-th qubit was $\ket{+}$,
and $\alpha_i=1$ otherwise.
Define the bit $\beta_i = 0$ if Bob's measurement outcome
for $i$-th qubit was $\ket{+}$,
and $\beta_i=1$ otherwise.
Define the bit $\gamma_i = 0$ if Charlie's measurement outcome
for $i$-th qubit was $\ket{+}$,
and $\gamma_i=1$ otherwise.
\end{enumerate}
\emph{For the simplicity of the presentation,
we will describe the procedure extracting the secret key
from $a_i$, $b_i$ and $c_i$.}
\begin{enumerate}
\setcounter{enumi}{6}
\item\label{l7} Alice chooses a subset $S \subset \{1$, \ldots, $2n\}$
with size $|S| = n$ uniformly randomly from subsets of 
$\{1$, \ldots, $2n\}$, and publicly announces the choice of $S$.
Alice, Bob and Charlie publicly announce $a_i$, $b_i$ and $c_i$
for $i\in S$ and compute the error rate
\[
q_1 = \max \left\{
\frac{|\{i\in S \mid a_i \neq b_i \}|}{|S|},
\frac{|\{i\in S \mid a_i \neq c_i \}|}{|S|}\right\}.
\]
\item\label{l8} Alice chooses a subset $S' \subset \{1$, \ldots, $2n'\}$
with size $|S'| = n'$ uniformly randomly from subsets of 
$\{1$, \ldots, $2n'\}$, and publicly announces the choice of $S'$.
Alice, Bob and Charlie publicly announce $\alpha_i$, $\beta_i$ and $\gamma_i$
for $i\in S'$ and compute the error rate
\begin{equation}
q_2 = 
\frac{|\{i \in S' \mid \alpha_i =\gamma_i \neq \beta_i \textbf{ or } \alpha_i =\beta_i \neq \gamma_i \}|}{|S'|}. \label{eq:q2}
\end{equation}
\end{enumerate}
Observe the difference between the definitions for $q_1$ and $q_2$.
\begin{enumerate}
\setcounter{enumi}{8}
\item\label{l9}
Alice, Bob and Charlie decide a linear code $C_1$ of length $n$
such that its decoding error probability is sufficiently small
over all the binary symmetric channel whose crossover probability
is close to $q_1$.
Let $H_1$ be a parity check matrix for $C_1$,
$\vec{a}$ be Alice's remaining (not announced) bits among $a_i$'s,
$\vec{b}$ be Bob's remaining bits among $b_i$'s,
and $\vec{c}$ be Charlie's remaining bits among $c_i$'s.
\item\label{l10}
Alice publicly announces the syndrome $H_1\vec{a}$.
\item\label{l11}
Bob compute the error vector $\vec{f}$ such that
$H_1\vec{f}=H_1\vec{b}-H_1\vec{a}$ by the decoding
algorithm for $C_1$. With high probability $\vec{b}-\vec{f} = \vec{a}$.
\item\label{l12}
Charlie compute the error vector $\vec{f'}$ such that
$H_1\vec{f'}=H_1\vec{c}-H_1\vec{a}$ by the decoding
algorithm for $C_1$. With high probability $\vec{c}-\vec{f'} = \vec{a}$.
\item\label{l13}
Alice chooses a subspace $C_2 \subset C_1$ with $\dim C_2 = nh(q_2)$
uniformly randomly, where $h$ denotes the binary entropy function,
and publicly announces her choice of $C_2$.
The final shared secret key is the coset $\vec{a}+C_2$.
\end{enumerate}

\section{Security proof and a lower bound on the key rate}
We shall show the unconditional security of our proposed
protocol by directly relating it to the quantum error correction by
the quantum CSS (Calderbank-Shor-Steane) codes \cite{calderbank96,steane96}.
To make this paper self-contained, we shall briefly review the
CSS code.

\subsection{Review of the CSS code}
For a binary vector
$\vec{v} = (v_1$, \ldots, $v_n) \in \mathbf{F}_2^n$,
where $\mathbf{F}_2$ is the Galois field with two elements,
we define the quantum state vector $\ket{\vec{v}}$ by
\[
\ket{\vec{v}} = \ket{v_1} \otimes \ket{v_2} \otimes \cdots \otimes
\ket{v_n}.
\]
For two binary linear codes $C_2 \subset C_1 \subset \mathbf{F}_2^n$,
the CSS code is the complex linear space spanned by the vectors
\[
\frac{1}{\sqrt{|C_2}|}
\sum_{\vec{w} \in C_2} \ket{\vec{v}+\vec{w}}, 
\]
for all $\vec{v}\in C_1$.
We also need parameterized CSS codes introduced in
Ref.~\cite{shor00}.
The parameterized CSS code for $\vec{x},\vec{z}\in\mathbf{F}_2^n$
is defined as the linear space spanned by
\[
\frac{1}{\sqrt{|C_2}|}
\sum_{\vec{w} \in C_2} (-1)^{(\vec{z},\vec{w})}\ket{\vec{x}+\vec{v}+\vec{w}}, 
\]
for all $\vec{v}\in C_1$,
where $(\cdot,\cdot)$ denotes the inner product.

For a linear code $C$ of length $n$,
define the linear code $CC$ of length $2n$ and dimension $\dim C$ by
\[
\{ \vec{c} \vec{c} \mid \vec{c} \in C\},
\]
where $\vec{c} \vec{c}$ is the concatenated vector of length $2n$.
In the security proof of our protocol,
we have to consider the CSS code defined by the pair
of linear codes $C_1C_1\supset C_2C_2$, whose orthonormal
basis is given by
\[
\frac{1}{\sqrt{|C_2}|}
\sum_{\vec{w} \in C_2} \ket{\vec{v}+\vec{w}}\ket{\vec{v}+\vec{w}}. 
\]
For vectors $\vec{x}$ and $\vec{z}\in \mathbf{F}_2^n$ of length
$n$, we define the parameterized CSS code as the complex linear
space spanned by
\begin{equation}
\frac{1}{\sqrt{|C_2}|}
\sum_{\vec{w} \in C_2} (-1)^{(\vec{z},\vec{w})}\ket{\vec{x}+\vec{v}+\vec{w}}\ket{\vec{x}+\vec{v}+\vec{w}}, \label{css4}
\end{equation}
for $\vec{v} \in C_1$.
Using the parameterized CSS code defined in Eq.~(\ref{css4}),
we shall show a security proof of our protocol.

\subsection{Security proof and analysis of the key rate}
We shall first show that our protocol is equivalent to
sending a parameterized CSS codeword with the parameter $\vec{z}$ randomly chosen.
If we fix $\vec{v}$ and $\vec{x}$ and choose $\vec{z}$ uniformly
randomly in Eq.~(\ref{css4}), then the resulting density operator
is
\begin{eqnarray}
&& \frac{1}{2^n|C_2|}\sum_{\vec{z}\in\mathbf{F}_2^n}
\left(
\sum_{\vec{w_1} \in C_2} (-1)^{(\vec{z},\vec{w_1})}\ket{\vec{x}+\vec{v}+\vec{w_1}}\ket{\vec{x}+\vec{v}+\vec{w_1}}
\right)\nonumber\\
&&
\left(
\sum_{\vec{w_2} \in C_2} (-1)^{(\vec{z},\vec{w_2})}\bra{\vec{x}+\vec{v}+\vec{w_2}}\bra{\vec{x}+\vec{v}+\vec{w_2}}\right)\nonumber\\
&=& \frac{1}{|C_2|}
\sum_{\vec{w}\in C_2} \ket{\vec{x}+\vec{v}+\vec{w}}\ket{\vec{x}+\vec{v}+\vec{w}}\bra{\vec{x}+\vec{v}+\vec{w}}\bra{\vec{x}+\vec{v}+\vec{w}}, \label{eq:mix0}
\end{eqnarray}
by the exactly same argument as Ref.~\cite{shor00}.

Denote the right hand side of Eq.~(\ref{eq:mix0})
by $\rho(\vec{x},\vec{v})$.
By a straightforward computation we can see
\begin{equation}
\frac{1}{4^n}
\sum_{\vec{x}\in\mathbf{F}_2^n}\sum_{\vec{v}\in\mathbf{F}_2^n}
\rho(\vec{x},\vec{v})
=
\frac{1}{2^n}\sum_{\vec{a}\in\mathbf{F}_2^n}\ket{\vec{a}\vec{a}}\bra{\vec{a}\vec{a}}.
\label{eq:mix}
\end{equation}
The right hand side of Eq.~(\ref{eq:mix})
means sending $\ket{00}$ or $\ket{11}$ $n$ times with equal probability,
which is exactly what Alice is doing in our protocol.
Announcing the syndrome $H_1\vec{a}$ in Step~\ref{l10}
is equivalent to announcing which $\vec{x}$ is chosen.

We shall consider the decoding of the imaginary transmission of
CSS codewords. A good review of decoding of CSS codes
is provided in Ref.~\cite{shor00}.
We regard Bob and Charlie as a single receiver.
The bit error correction is actually performed in our protocol,
therefore we have to ensure that it can be separately executed
by Bob and Charlie.
On the other hand, the phase error correction is not actually
performed, so it has not to be separately executable.

Observe that the bit error correction for the leftmost $n$
qubits in the CSS code
defined in Eq.~(\ref{css4}) can be done separately from
the rightmost $n$ qubits in Eq.~(\ref{css4}),
which means that Bob and Charlie can correct their bit flip errors
with their local operations and that they do not need their cooperation
for bit error correction.

We shall consider the phase error correction.
Let $Z_i$ be the phase error occurred at $i$-th qubit
in Eq.~(\ref{css4}). We can see that
$Z_i$ and $Z_{n+i}$ have the same effect on the quantum codeword
in Eq.~(\ref{css4}), and that $Z_i \otimes Z_{n+i}$ does
\emph{not} change the quantum codeword in Eq.~(\ref{css4}).

Let
\[
H_2 = \left(
\begin{array}{c}
\vec{h}_1\\
\vdots\\
\vec{h}_{n-\dim C_2}
\end{array}
\right)
\]
be a parity check matrix for $C_2^\perp$,
then
\[
H'_2 = \left(
\begin{array}{c}
\vec{h}_1\vec{h}_1\\
\vdots\\
\vec{h}_{n-\dim C_2}\vec{h}_{n-\dim C_2}
\end{array}
\right)
\]
is a parity check matrix for $(C_2C_2)^\perp$.

Suppose that a phase error
\begin{equation}
Z^{z_1}\otimes \cdots \otimes Z^{z_{2n}} \label{errorz1}
\end{equation}
occurred with the quantum state (\ref{css4}),
where $z_i \in \mathbf{F}_2$ for $i=1$, \ldots, $2n$.
Let $\vec{e} = (z_1$, \ldots, $z_{2n}) \in \mathbf{F}_2^{2n}$.
{From} the measurement outcomes in the phase error correction
process, we can know
\begin{equation}
(\vec{h}_i \vec{h}_i, \vec{e})\label{eq:synd}
\end{equation}
for $i=1$, \ldots, $n-\dim C_2$.
Observe that the error (\ref{errorz1}) and
\[
Z^{z'_1}\otimes \cdots \otimes Z^{z'_{2n}}
\]
has the same effect on the state (\ref{css4}) if
$z_i + z_{n+i} = z'_i + z'_{n+i}$ for all $i$, where the addition is
considered in $\mathbf{F}_2$.
Thus, in order to correct phase errors,
we have to find the binary vector
\begin{equation}
\vec{e'} = (e_1 + e_{n+i}, \ldots, e_n+e_{2n})
\end{equation}
from the data given in Eq.~(\ref{eq:synd}).
Also observe that
\[
(\vec{h}_i \vec{h}_i, \vec{e}) = (\vec{h}_i, \vec{e'}).
\]
Finding $\vec{e'}$ from $(\vec{h}_i, \vec{e'})$ is exactly 
the ordinary decoding process for the classical linear code $C_2^\perp$.
Also observe that the probability of $e_i + e_{n+i} = 1$ is given
by $q_2$ in Eq.~(\ref{eq:q2}) in our situation.

It is proved in Ref.~\cite{watanabe06} that
random choice of [$n-h(q_2)$]-dimensional subspace $C_2$ in $C_1$
almost always gives the low phase error decoding probability.
This fact is also stated without proof in Ref.~\cite{shor00}.
If the choice of $C_1$ is appropriate,
then the fidelity of quantum error correction in the
imaginary transmission of the CSS codeword (\ref{css4})
is close to $1$,
which implies that the eavesdropper Eve can obtain little
information by the same argument as Ref.~\cite{hamada-qkd},
which shows the security of the BB84 protocol
directly relating it to the quantum error correction
without use of entanglement distillation argument.

It was shown in Corollary 2 of Ref.~\cite{csiszar82}
that there exists a linear code $C_1$ of information rate
$1-h(q_1)$ satisfying the condition in Step~\ref{l9}.
Therefore, we can extract $1-h(q_1)-h(q_2)$ bit of secret key
from one bit of the raw bits $\vec{a}$.

\section{Conclusion and Discussion}
We have proposed a protocol 
that allows Alice, Bob and Charlie to share a common secret key
at once. This is the first such protocol without use of entangled
states.
However, the amount of extracted common secret key
per single photon transmission is lower than the repeated use
of the BB84 protocol and the well-known
postprocessing described in Ref.~\cite{singh03}.
Finding a multiparty QKD protocol with higher efficiency is
a future research agenda.

\section*{Acknowledgment}
The author would like to thank Shun Watanabe and
Tomohiko Uyematsu for helpful discussions.
This research was partly supported
by the Japan Society for the Promotion of Science
under Grants-in-Aid for Young Scientists No.\ 18760266.


\end{document}